\documentclass[aps,prx,reprint]{revtex4-2}
\usepackage{bm}
\usepackage{graphicx}
\usepackage{subfigure}
\usepackage{amsmath}
\usepackage{amssymb}
\usepackage{color}
\usepackage[a4paper,left=1.5cm, right=1.3cm, top=3.0cm,bottom=2cm]{geometry}

\usepackage{comment}

\begin{document}
\title{Supergranule aggregation for constant heat flux-driven turbulent convection}
\author{Philipp P. Vieweg}
\affiliation{Institut f\"ur Thermo- und Fluiddynamik, Technische Universit\"at Ilmenau, Postfach 100565, D-98684 Ilmenau, Germany}
\author{Janet D. Scheel}
\affiliation{Department of Physics, Occidental College, 1600 Campus Road M21, Los Angeles, CA  90041, USA}
\author{J\"org Schumacher}
\affiliation{Institut f\"ur Thermo- und Fluiddynamik, Technische Universit\"at Ilmenau, Postfach 100565, D-98684 Ilmenau, Germany}
\affiliation{Tandon School of Engineering, New York University, New York City, NY 11201, USA}
\date{\today}

\begin{abstract}
Turbulent convection processes in nature are often found to be organized in a hierarchy of plume structures and flow patterns. The gradual aggregation of convection cells or granules to a supergranule which eventually fills the whole horizontal layer is reported and analysed in spectral element direct numerical simulations of three-dimensional turbulent Rayleigh-B\'{e}nard convection at an aspect ratio of $60$. The formation proceeds over a time span of more than $10^4$ convective time units for the largest accessible Rayleigh number and occurs only when the turbulence is driven by a constant heat flux which is imposed at the bottom and top planes enclosing the convection  layer. The resulting gradual inverse cascade process is observed for both temperature variance and turbulent kinetic energy. An additional analysis of the leading Lyapunov vector field for the full turbulent flow trajectory in its high-dimensional phase space demonstrates that turbulent flow modes at a certain scale continue to give rise locally to modes with longer wavelength in the turbulent case. As a consequence successively larger convection patterns grow until the horizontal extension of the layer is reached. This instability mechanism, which is known to exist near the onset of constant heat flux-driven convection, is shown here to persist into the fully developed turbulent flow regime thus connecting weakly nonlinear pattern formation with the one in fully developed turbulence. We discuss possible implications of our study for observed, but not yet consistently numerically reproducible,  solar supergranulation which could lead to improved simulation models of surface convection in the Sun.   
\end{abstract}
\keywords{Rayleigh-B\'{e}nard convection}
\maketitle

\section{Introduction}
Turbulent convection, the essential mechanism by which heat is transported in natural flows, manifests often in a hierarchy of structures and flow patterns. Clusters of clouds over the warm oceans in the tropics on Earth \cite{Mapes1993} or giant storm systems in the atmospheres of the big gas planets Jupiter \cite{Young2017} and Saturn \cite{Garcia2013} illustrate this phenomenon. One of the most prominent astrophysical examples is the convection zone in the outer 30\% of the Sun \cite{Schumacher2020}. Convection cells are termed granules if they have an extension of $\ell_{\rm G}\sim 10^3$ km and a lifetime of about 10 minutes. These granules form the basic pattern at the solar surface where a heat flux drives convection \cite{Frenkiel1952,Riethmueller2014}. Spectral observations reveal supergranules with extensions of $\ell_{\rm SG}\sim 30 \ell_{\rm G}$ and a lifetime of a day as the next larger building block in this hierarchy, detected either by line shifts in optical observations \cite{Leighton1962}, helioseismology \cite{Dalsgaard2002} or granule tracking \cite{Rincon2018}. Between the granule and supergranule scale a whole range of mesoscales exists, but without an additional prominent scale. Giant cells that extend across the whole convection zone could be a third stage in this hierarchy \cite{Hathaway2013,Brandenburg2016,Anders2019}, but this is still an open question. Different physical effects have  been proposed for the formation of supergranules, such as helium recombination in the upper convection zone \cite{Lord2014}, self-organisation of granules \cite{Rincon2018}, or dynamical constraints by deeper convection at scales $\ell \gg \ell_{\rm SG}$ that are affected by the slow rotation of the Sun \cite{Featherstone2016}. Numerical simulations that model convection and try to predict the spectral measurements have still been unable to develop supergranules self-consistently \cite{Hanasoge2014,Cossette2016}. 

Here, we demonstrate the aggregation of granules to a large-scale supergranule in the simplest setting of convection without additional physical processes such as radiation, rotation or magnetic fields involved in heat and momentum transfer. This turbulent Rayleigh-B\'{e}nard convection (RBC) case in the Boussinesq limit is often considered as the paradigm for convective turbulence with its many facets \cite{Ahlers2009,Chilla2012}. Our three-dimensional direct numerical simulations (DNS) differ in three important ways from the majority of numerical studies in RBC: (1)  they are subject to constant heat flux boundary conditions at the top and bottom; (2) they require simulations that are run on the order of $10^4$ convective free-fall time units and even more; and (3) they are conducted in sufficiently extended layers. Layers with a fixed aspect ratio $\Gamma=L/H=60$ with the horizontal length $L$ and the layer height $H$ are considered in this paper. We observe the gradual supergranule formation for all accessible Rayleigh numbers up to ${\rm Ra}\approx 7.7 \times 10^7$, a dimensionless measure for the vigor of convective turbulence. Their formation proceeds despite the fact that the flow becomes fully time-dependent and turbulent. This is in a regime for which one would not expect a pattern coherence across the whole domain, given that ${\rm Ra}$ is far beyond the critical Rayleigh number ${\rm Ra}_c$ for the onset of the primary linear flow instability \cite{Sparrow1964,Hurle1967} or subsequent secondary instabilities of the onset pattern \cite{Busse1967,Chapman1980}. 

We confirm the continued gradual aggregation trend into the fully turbulent regime by the Lyapunov vector field of the largest Lyapunov exponent \cite{Pikovsky2016} of the turbulent states (see refs. \cite{Egolf2000,Scheel2006,Levanger2019} for similar analyses to characterize pattern defects in weakly nonlinear convection). The Lyapunov vector analysis probes here the growth of linear instabilities and of the corresponding scales of our high-dimensional nonlinear dynamical system. In particular, we show that the leading Lyapunov vector field becomes coarser as time proceeds which suggests that the turbulent flow remains unstable at a given scale with respect to longer-wavelength instabilities until the domain size has been reached. Indeed such an ongoing inverse cascade of energy and thermal variance can be clearly shown by the power spectra. The supergranule becomes better visible in the velocity and temperature once a time-windowed averaging is applied that suppresses the turbulent fluctuations and the faster converging and diverging flows in the small-scale convective granules. We find supergranules  independent of the boundary conditions of the velocity field. 

Furthermore, we show that the supergranule is absent in DNS with constant temperature boundary conditions at the top and bottom planes of the RBC layer. For these cases, the formation of the recently comprehensively investigated turbulent superstructures \cite{Hartlep2003,Parodi2004,Hartlep2005,Hardenberg2008,Bailon2010,Emran2015,Stevens2018,Pandey2018,Fonda2019,Green2020,Krug2020} -- well-ordered patterns of temperature and velocity with characteristic convection roll widths up to $\Lambda/2 \sim 3-4H$ \cite{Stevens2018,Pandey2018,Krug2020} -- takes place. $\Lambda$ is the characteristic scale or wavelength of the pattern. Big velocity field condensates have been studied in two-dimensional \cite{Smith1993} and quasi-two-dimensional fluid turbulence \cite{Smith1996,Celani2010,Musacchio2019} to analyse the dependence of the inverse cascade on the energy injection scale. These settings are different to RBC where the driving of the fluid motion proceeds by thermal plumes that have a typical width of the order of the thermal boundary layer thickness $\delta_T \ll H\ll L$. A slowly progressing clustering of thermal plumes in RBC has been studied in von Hardenberg et al. \cite{Hardenberg2008} for $L\le 12\pi H$ reaching roll widths of $\Lambda/2 \sim 3H$ for similar Rayleigh and Prandtl numbers. The generation of a large-scale anisotropy in turbulent convection for free-slip velocity conditions at the walls requires additional rotation about a horizontal axis in the three-dimensional case as shown in ref. \cite{Hardenberg2015}.

Hurle et al. \cite{Hurle1967} studied the linear stability of an infinitely extended two-dimensional thermal convection layer at rest for the constant flux case analytically. They detected a critical wavenumber $k_c=0$ and a critical Rayleigh number ${\rm Ra}_c = 6 ! = 720$ for no-slip velocity conditions and ${\rm Ra}_c = 5 ! = 120$ for free-slip conditions. This implies that the pair of counterrotating convection rolls at the onset of convection will always extend to the largest possible wavelength $\Lambda=L< \infty$ in a finite cell. Instabilities of finite-amplitude convection rolls for Rayleigh numbers slightly above ${\rm Ra}_c$ showed that each mode is unstable to one longer wavelength \cite{Chapman1980}. Interestingly, this gradual aggregation process has not been observed in previous turbulent RBC simulations with constant flux boundary conditions \cite{Verzicco2004,Verzicco2008,Doering2009}, most probably because they were conducted in smaller aspect ratio domains and for shorter total integration times. 

Our study suggests that the mechanisms of supergranule formation in a simple convection flow are related to linear instabilities in the turbulent flow that give rise to longer-wavelength structures. Even though the RBC flow operates at Rayleigh numbers that are up to nearly 5 orders of magnitude above the critical Rayleigh number for the onset of the primary instability, a cell with the longest wavelength is still formed without any additional physical mechanism at work in the unstably stratified layer. Our investigation can thus shed a new light on the fundamentals of solar granulation processes.
%------------------------------------------------------------------------------------------
\begin{table*}[t]
\centering
\begin{tabular}{l r r l r r r r r}
\hline\hline
\multicolumn{1}{l}{Run}	& \multicolumn{1}{c}{${\rm Ra}$} & \multicolumn{1}{l}{${\rm Ra}/{\rm Ra}_c$}	& \multicolumn{1}{l}{${\rm Pr}\;\;$}	& \multicolumn{1}{c}{$N_{e}$}	& \multicolumn{1}{l}{$N$}	& \multicolumn{1}{c}{$\tilde{t}_{r}$}	& \multicolumn{1}{c}{${\rm Nu}$}	& \multicolumn{1}{c}{${\rm Re}$}\\
\hline
Nfs1	&$10,430$ 		&87			& 1		& 160,000 		& 7 	& 4,000		& $3.93 \pm 0.12$	& $26.4 \pm 0.4$\\
Nfs2	&$203,600$ 		&1697		& 1		& 160,000		& 11 	& 6,500		& $6.74 \pm 0.10$	& $81.4 \pm 0.7$\\
Nfs3	&$3,928,000$ 	&32733		& 1		& 1,280,000 	& 7		& 10,000	& $12.30 \pm 0.16$	& $229.0 \pm 1.4$\\
Nfs4	&$76,890,000$ 	&640750		& 1		& 11,022,400 	& 7   	& 19,000	& $23.50 \pm 0.24$	& $635.9 \pm 3.1$\\
Dfs2  	&$38,500$		&58			& 1		& 160,000 		& 11	& 1,450 	& $5.29 \pm 0.04$	& $74.2 \pm 0.2$\\
Dfs3  	&$385,000$ 		&580		& 1		& 1,280,000		& 7		& 1,100 	& $10.21\pm 0.04$	& $215.8\pm 0.5$\\
\hline\hline
\end{tabular}
\caption{Parameters of the direct numerical simulation runs. We list the Rayleigh number ${\rm Ra}$, the ratio to the critical Rayleigh number ${\rm Ra}_c$, the Prandtl number ${\rm Pr}$, the total number of spectral elements in the simulation domain $N_{e}$, the polynomial order $N$ on each spectral element, the total dimensionless runtime of the simulation $\tilde{t}_{r}$ in units of the corresponding free-fall times $t_{f}$, the resulting Nusselt number ${\rm Nu}$, and the Reynolds number ${\rm Re}$. All values correspond to the late state of the flow where the supergranule is completely established for the Neumann cases. ${\rm Nu}$ and ${\rm Re}$ are determined from $50$ snapshots within the last $500 t_{f}$ of each simulation. Error bars are determined by the standard deviation.}
\end{table*}
%------------------------------------------------------------------------------------------

\section{Numerical analysis}
\label{sec:method_data}

We consider here the simplest turbulent convection configuration, the three-dimensional Boussinesq case which couples the temperature field $T({\bm x},t)$ and the velocity vector field ${\bm u}({\bm x},t)$ in an incompressible fluid \cite{Ahlers2009,Chilla2012} with ${\bm u}=(u_x, u_y, u_z)$ and ${\bm x}=(x, y, z)$. In this case the mass density is a linear function of the temperature deviation from the reference value. The Cartesian domain $V=L\times L\times H$ applies periodic boundary conditions in both lateral directions, $x$ and $y$ for all fields.
Regarding the vertical direction, the following boundary conditions are applied at the bottom and top plates at $z=0,H$. For the velocity field, these are either no-slip (ns) or free-slip or stress-free (fs) conditions. They are given by
\begin{align}
\text{(ns)} & \quad u_x=u_y=u_z=0\,,\\
\text{(fs)} & \quad \frac{\partial u_x}{\partial z}= \frac{\partial u_y}{\partial z}=0 \quad \text{and} \quad u_z=0\,.
\end{align}
Thermal conditions are either of Dirichlet (D) or Neumann (N) type,
\begin{align}
\text{(D)} & \quad T(z=0)=T_{\rm bot} \quad \text{and} \quad T(z=H)=T_{\rm top}\,,\\
\text{(N)} & \quad \frac{\partial T}{\partial z}\Bigg|_{z=0}=\frac{\partial T}{\partial z}\Bigg|_{z=H}=-\beta\,,
\end{align}
with $\beta>0$. We adopt as units of length and time the layer height $H$ and the free-fall time $t_{f} = H / U_f$ with the free-fall velocity $U_f$ to rescale the equations in a dimensionless form. The latter is defined by $U_f = \sqrt{\alpha g (T_{\rm bot}-T_{\rm top})H}$ for the Dirichlet (D) case with the characteristic temperature that is given by the difference $\Delta T=T_{\rm bot}-T_{\rm top}>0$. The quantity $\alpha$ is the isobaric expansion coefficient and $g$ is the acceleration due to gravity. In case of Neumann (N) boundary conditions, one obtains  $U_f=\sqrt{\alpha g \beta H^2}$ while the characteristic temperature is $\beta H$. The dimensionless equations of motion follow as
\begin{align}
\tilde{\nabla} \cdot \tilde{\bm{u}} & = 0\,, \label{ce}\\
\frac{\partial \tilde{\bm{u}}}{\partial \tilde{t}} + ( \tilde{\bm{u}} \cdot \tilde{\nabla} ) \tilde{\bm{u}} & = - \tilde{\nabla} \tilde{p} + \sqrt{\frac{\rm Pr}{{\rm Ra}_{D,N}}} \tilde{\nabla}^{2} \tilde{\bm{u}} + \tilde{T} \bm{e}_{z} \,,\\
\frac{\partial \tilde{T}}{\partial \tilde{t}} + ( \tilde{\bm{u}} \cdot \tilde{\nabla} ) \tilde{T} & = \frac{1}{\sqrt{{\rm Ra}_{D,N} {\rm Pr}}} \tilde{\nabla}^{2} \tilde{T} \label{te}\,,
\end{align}
with the dimensionless pressure field $\tilde{p}$. Dimensionless quantities are indicated by a tilde in the equations. The Prandtl and Rayleigh numbers are given by
\begin{equation}
{\rm Pr}=\frac{\nu}{\kappa}\,,\;\;
{\rm Ra}_D=\frac{g\alpha\Delta T H^3}{\nu\kappa}\;\;\mbox{and}\;\;
{\rm Ra}_N=\frac{g\alpha \beta H^4}{\nu\kappa} \,,
\end{equation}
with the kinematic viscosity of the fluid $\nu$ and the temperature diffusivity $\kappa$. 

The equations of motion are solved numerically with the spectral element method nek5000 \cite{Fischer1997,Scheel2013}. The polynomial order $N$ on each element and the total spectral element number $N_e$ are chosen properly such that the steep gradients near the top and bottom walls and the Kolmogorov scale $\eta_K$ can be resolved sufficiently (see \cite{Scheel2013} for more details). We varied the order $N$ at fixed Rayleigh and Prandtl numbers to verify that the supergranule formation, the mean profiles of temperature, and the temperature variance spectra are unaffected (see appendix A for further details on the resolution tests).  

The turbulent heat transfer across the convection layer is determined by the dimensionless Nusselt number which is given for constant flux boundary conditions (N) by (see also  ref. \cite{Otero2002} for a derivation)
\begin{equation}
{\rm Nu}_{N} =\frac{1}{\langle \tilde{T}(\tilde{z}=0)\rangle_{\tilde{A},\tilde{t}}-\langle \tilde{T}(\tilde{z}=1)\rangle_{\tilde{A},\tilde{t}}}\,,
\label{NuN}
\end{equation}
and for the constant temperature conditions (D) by 
\begin{equation}
{\rm Nu}_{D} =  - \Bigg\langle \frac{\partial \tilde{T}}{\partial \tilde{z}}\Bigg |_{\tilde{z}=0}\Bigg\rangle_{\tilde{A},\tilde{t}} = - \Bigg\langle \frac{\partial \tilde{T}}{\partial \tilde{z}}\Bigg |_{\tilde{z}=1}\Bigg\rangle_{\tilde{A},\tilde{t}}\,.
\label{NuD}
\end{equation}
The symbol $\langle\cdot\rangle_{\tilde{A},\tilde{t}}$ denotes a combined average over the horizontal cross section $\tilde{A}=\Gamma^2$ with the aspect ratio $\Gamma=L/H$ and time $\tilde{t}$. In comparison the global turbulent momentum transport is quantified by the Reynolds number, which is for numerical studies defined as
\begin{equation}
{\rm Re} = \sqrt{\frac{{\rm Ra}_{D,N}}{\rm Pr} \thickspace \langle \tilde{\bm{u}}^{2} \rangle_{\tilde{V}, \tilde{t}}} .
\end{equation}

The boundary conditions (ns), (fs) for the velocity field and (D), (N) for the temperature field can be combined to four different sets of runs at several Rayleigh numbers. All four groups of boundary conditions were investigated. The main focus of our presentation will be on the series Nfs1 to Nfs4 which is listed in Table 1. This combination of boundary conditions comes closest to the solar convection case, that motivates our study. Note also that the Rayleigh numbers for cases Dfs and Nfs are related to each other by
%------------------------------------------------------------------------------------------
\begin{equation}
{\rm Ra}_N={\rm Nu}_D{\rm Ra}_D\,.
\label{comparison}
\end{equation} 
%------------------------------------------------------------------------------------------
Since the runs with Dirichlet conditions served always as the starting point, the values of ${\rm Ra}_N$ follow as given in Table 1. 

In the remaining manuscript, we will drop the tildes on all dimensionless quantities. The presentation of the results is continued in dimensionless units, for example $0 \le z \le 1$. Note also that our choice of the characteristic temperature in the Neumann case can cause values of $T$ smaller and larger than $[0,1]$, which again lead to a global mean $\langle T\rangle_{V,t}=1/2$ as in case D. Here, we do not rescale these temperatures since we do not directly compare the temperature statistics between cases D and N.

%------------------------------------------------------------------------------------------
\begin{figure*}[t]
\begin{center}
\includegraphics[scale=1.00]{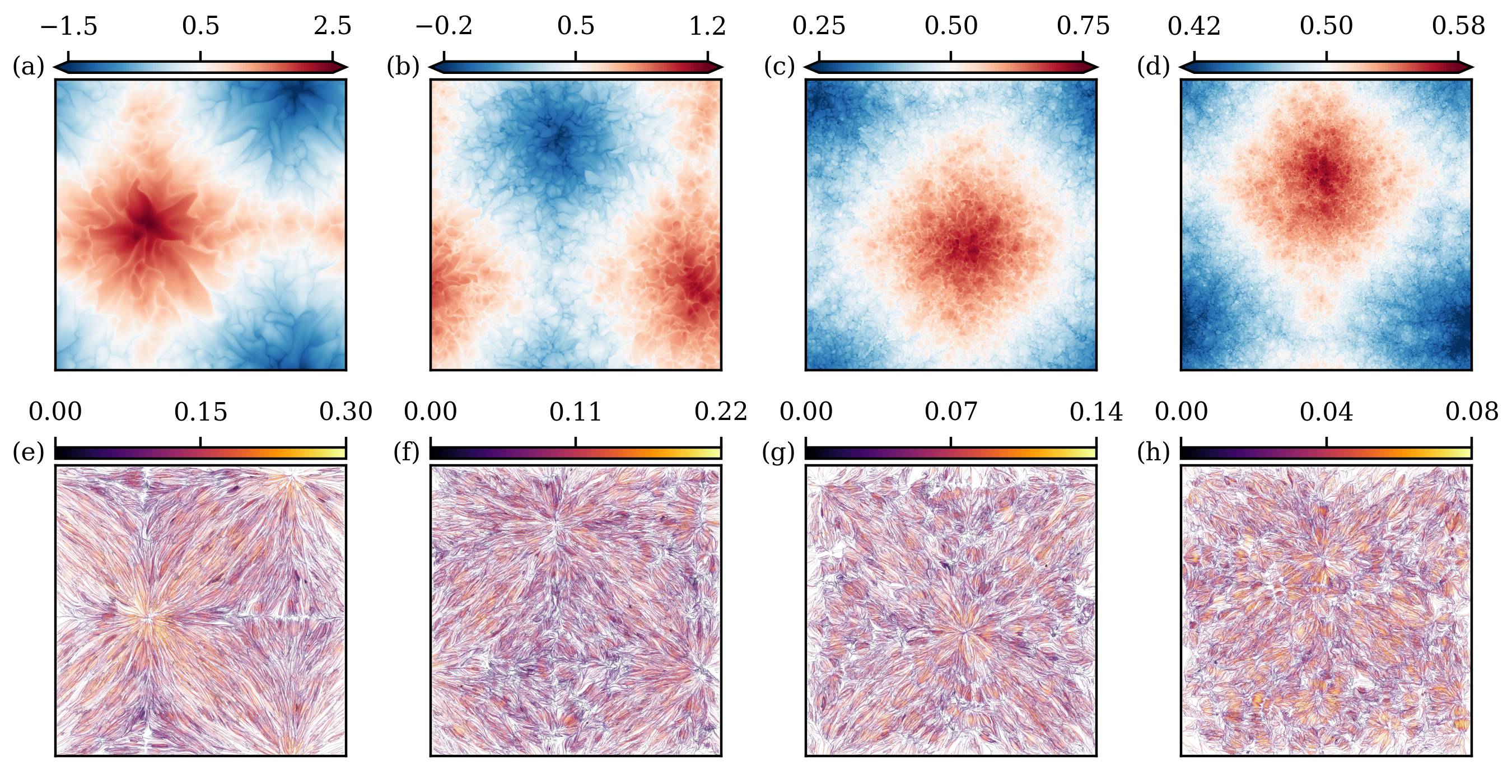}
\caption{Supergranule structure in the convection layer for different Rayleigh numbers ${\rm Ra}_N$. The top row displays instantaneous snapshots of the temperature field at $z_0=1-\delta_T/2$ where the dimensionless thermal boundary layer thickness is given by $\delta_T=1/(2 {\rm Nu}_N)$. Snapshots were taken at $t=4,010$ in (a,e), $6,400$ in (b,f), $10,000$ in (c,g), and $19,000$ in (d,h). The bottom row shows time-averaged plots of the streamlines viewed from the top. The color corresponds to the velocity magnitude. The averaging time interval of $500 t_f$ is always taken in the final phase of the simulation. Data are for Nfs1 in (a,e), Nfs2 in (b,f), Nfs3 in (c,g), and Nfs4 in panels (d,h) as listed in Table 1. All fields are displayed for the whole cross-section of size $L \times L = 60 \times 60$.}
\label{fig1}
\end{center}
\end{figure*} 
%------------------------------------------------------------------------------------------
\begin{figure}
\begin{center}
\includegraphics[scale=1.00]{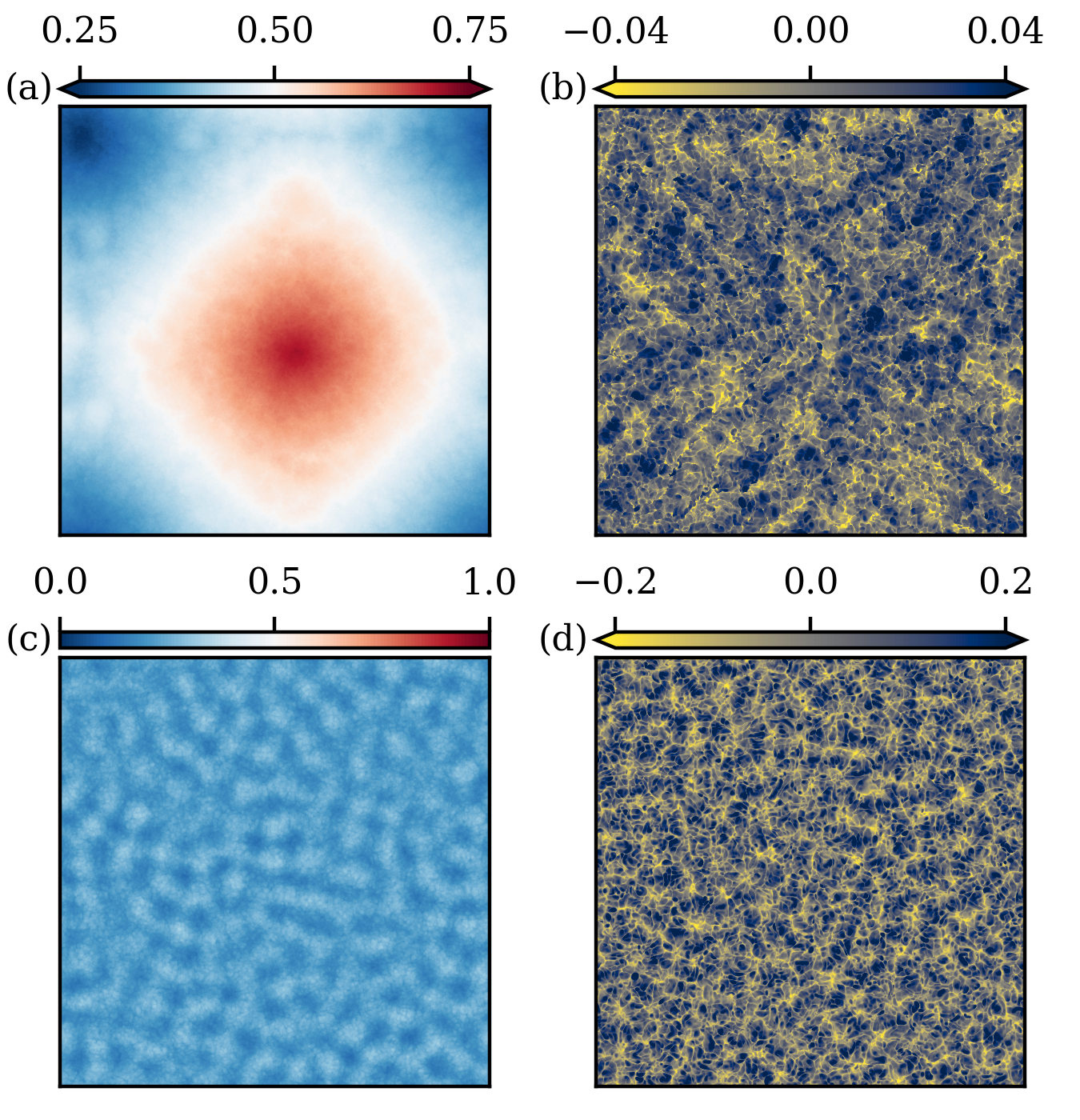}
\caption{Comparison of Neumann and Dirichlet cases. The left column shows the time-averaged temperature fields $\langle T(x,y)\rangle_t$ at $z_0=1-\delta_T/2$ while the right column plots the temperature fluctuations $\theta(x,y)$ at a given instant and the same height. The Rayleigh numbers are ${\rm Ra}_N={\rm Nu}_D {\rm Ra}_D=3.93\times 10^6$ (case Nfs3) in panels (a,b) and ${\rm Ra}_D=3.85\times 10^5$ (case Dfs3) in panels (c,d). The whole cross-section of size $L \times L = 60 \times 60$ is displayed.}
\label{fig2}
\end{center}
\end{figure}
%------------------------------------------------------------------------------------------

\section{Supergranule formation}
%------------------------------------------------------------------------------------------
\begin{figure*}
\begin{center}
\includegraphics[scale=1.00]{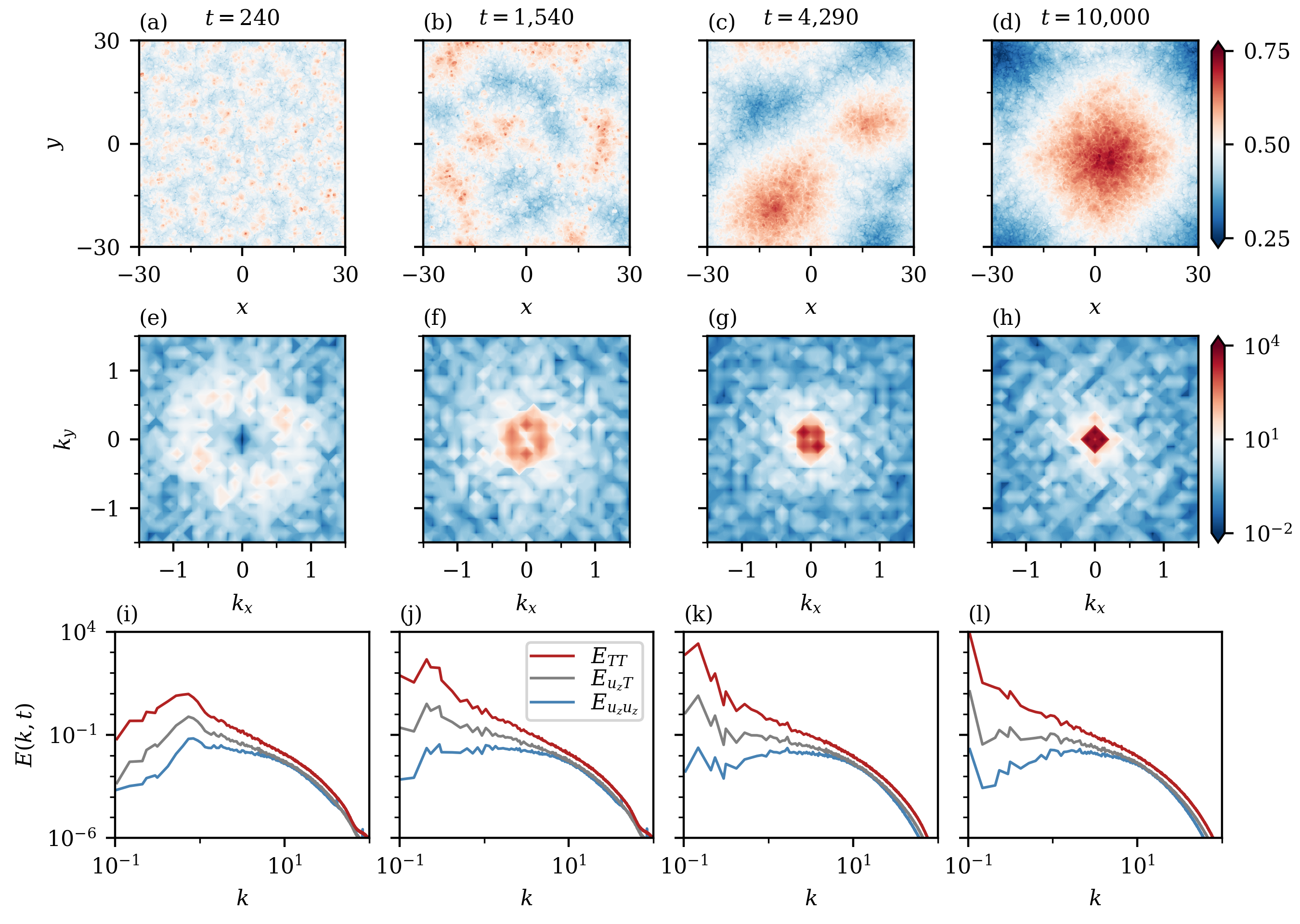}
\caption{Gradual aggregation in wavevector space. The process is shown for the supergranule that forms at ${\rm Ra}_N=3.93\times 10^6$ for a time period of $10^4$ free-fall times as an example. (a--d) Temperature contour plots close to the top plane at $z_0=1-\delta_T/2$. Times for the snapshots are indicated. The color bar applies for all panels in this row. (e--h) Corresponding Fourier spectra taken from temperature fluctuations at the same horizontal plane with respect to the horizontal wave vector components $(k_x, k_y)$. The wavenumber range on both axes magnifies the smallest value to demonstrate the variance accumulation in the largest possible modes. The color bar is again the same for all panels in this row. (i--l) Additionally azimuthally averaged Fourier spectra with respect to $k=(k_x^2+k_y^2)^{1/2}$. The Fourier transform of three different quantities is displayed as indicated in the legend in panel (j): temperature variance spectrum, co-spectrum of the turbulent convective heat flux, and the kinetic energy spectrum with respect to the vertical velocity component.}
\label{fig3}
\end{center}
\end{figure*}
%------------------------------------------------------------------------------------------

Figure \ref{fig1} illustrates the final stages of the simulations Nfs1 to Nfs4 at four different Rayleigh numbers between $10^4\lesssim {\rm Ra}_N\lesssim 10^8$ (see table 1). The top row shows the temperature contour snapshots taken in the final stages of the simulations close to the upper surface of the layer. A pair of large square-shaped convection cells, which we term supergranules, are observed in all 4 cases. Due to the periodic boundary conditions in $x$ and $y$, they are partly distributed across the lateral boundaries of the domain. These structures, which are expected at and slightly above onset of convection in this setting \cite{Chapman1980}, thus continue to exist into the fully developed turbulent regime. They are clearly visible in all cases as a hotter and colder background structure of the temperature field. Superposed is a fine-scale granule pattern that would also be observable for other RBC cases. It is related to the instability of fragments of the thermal boundary layer and the related thermal plume formation. The bottom row displays the corresponding surface velocity streamlines, again viewed from the top. They are obtained after a time-averaging over $500$ convective time units centered around the temperature plots. 
%\change{We provide a video of the gradual aggregation in run Nfs4 as Supplementary Material.}

Figure \ref{fig2} underlines the different character of the large-scale patterns of the Dirichlet and Neumann cases at the related Rayleigh numbers ${\rm Ra} \sim 10^5-10^6$. We also decompose the temperature into $T({\bm x},t)=\langle T({\bm x})\rangle_t+\theta({\bm x},t)$ where $\langle\cdot\rangle_t$ stands for a time average over an interval of $500$ free-fall times $t_f$. The horizontal cuts which are always taken at the height $z_0=1-\delta_T/2$ decompose the data of the Dirichlet case into a time-averaged superstructure pattern with a characteristic wavelength that agrees with those in refs. \cite{Stevens2018,Pandey2018} and a fine skeleton of plume-ridges. This is in contrast to the Neumann case, where the supergranule is now clearly revealed. Interestingly, the instantaneous temperature fluctuation patterns for the Neumann case are more similar to those of the corresponding Dirichlet case. In appendix B we demonstrate furthermore that our observation is not altered by a substitution of free-slip with no-slip conditions which is consistent with the behaviour at the onset of convection \cite{Hurle1967}.  

Figure~\ref{fig3} quantifies the gradual formation of the supergranule by quantities in the physical and Fourier spectral space for the case Nfs3 at ${\rm Ra}_N=3.93\times 10^6$. We note here that all simulations started with a perturbation of the linear diffusive equilibrium profile with random noise where the fluid is at rest. The very initial evolution where the flow relaxes into a statistically stationary regime for the fluctuations of the turbulent fields is discarded. In the course of the nonlinear evolution, a growth of the scale of the convection patterns, here the temperature field close to top plate, is observed in panels (a--d) which shows similarities to a phase separation process that has been analysed in binary-fluid mixtures \cite{Perlekar2014,Perlekar2017}. The slow aggregation proceeds over $\sim 10^4$ free-fall times; we find that the time this large-scale structure formation takes grows with respect to Rayleigh number (see also Fig. \ref{fig1}). This time span
is longer than a vertical diffusion time scale $\tau_v=\sqrt{\rm Pr\,Ra}$, but significantly shorter than a horizontal diffusion scale $\tau_h=\Gamma^2\tau_v$. 

Panels (e--h) plot the squared magnitude of the two-dimensional Fourier transforms with respect to the horizontal coordinates $x$, $y$ of the temperature. We display $|\hat T(k_x,k_y,z_0,t_0)|^2$, in logarithmically increasing contour levels at four different times $t_0$ for $z_0=1-\delta_T/2$. The data correspond to those in the top row of the figure. We observe the slow transformation from the ring-like maximum, which is also observed in the Dirichlet case \cite{Hartlep2003,Pandey2018,Krug2020}, to a condensate in the four (next neighbor) discrete planar wavevectors ${\bm k}_{1,2}=(\pm k_{\rm min},0)$ and ${\bm k}_{3,4}=(0,\pm k_{\rm min})$ around the horizontally homogeneous mode with ${\bm k}=0$ (see panel h) which cannot be accessed in a domain with a finite box or periodicity of length $L$. The magnitudes of these 4 wavevectors correspond to the largest wavelength a convection structure can take in a domain, namely $\Lambda=\Gamma$. Here $k_{\rm min}=2\pi/\Gamma \approx 0.1$ and thus each of the two supergranules has an approximate size of $\Lambda/2\approx 30$.  This is different from an infinitely extended domain which would result in a continuum of possible wavevectors and a critical wavenumber $k_c=0$. Thus we interpret the accumulation of thermal variance and kinetic energy in ${\bm k}_{1, 2, 3, 4}$ as the finite-size relic of the primary linear instability mechanism which is not forgotten by the system. We have verified that a similar, but faster, aggregation takes place in boxes at smaller aspect ratio and the behavior is the same when the analysis is repeated in the midplane. 

The panels in the bottom row of Fig. \ref{fig3} show the azimuthally averaged spectra for different quantities: (i) temperature $E_{TT}(k,z=z_0)$ as in top and middle row, (ii) vertical velocity component $E_{u_zu_z}(k,z=z_0)$, and (iii) convective heat flux $E_{u_z T}(k,z=z_0)$. For all temperature fields, that enter the spectral analysis, the area mean $\langle T(z_0)\rangle_{A,t}$ is subtracted. Spectra are given by 
%------------------------------------------------------------------------------------------
\begin{equation}
E_{uv}(k,t)=\frac{1}{2\pi} \int_0^{2\pi} \Re[\hat{u}(k,\phi,t) \hat{v}^{\ast}(k,\phi,t)] d\phi\,.
\label{comparison1}
\end{equation} 
%------------------------------------------------------------------------------------------
for $uv=\{TT, u_zu_z, u_zT\}$. All quantities display clearly an accumulation of spectral density at the lowest wavenumber suggesting an inverse cascade process that leads to the formation of the supergranule.
%------------------------------------------------------------------------------------------
\begin{figure}
\begin{center}
\includegraphics[scale=1.00]{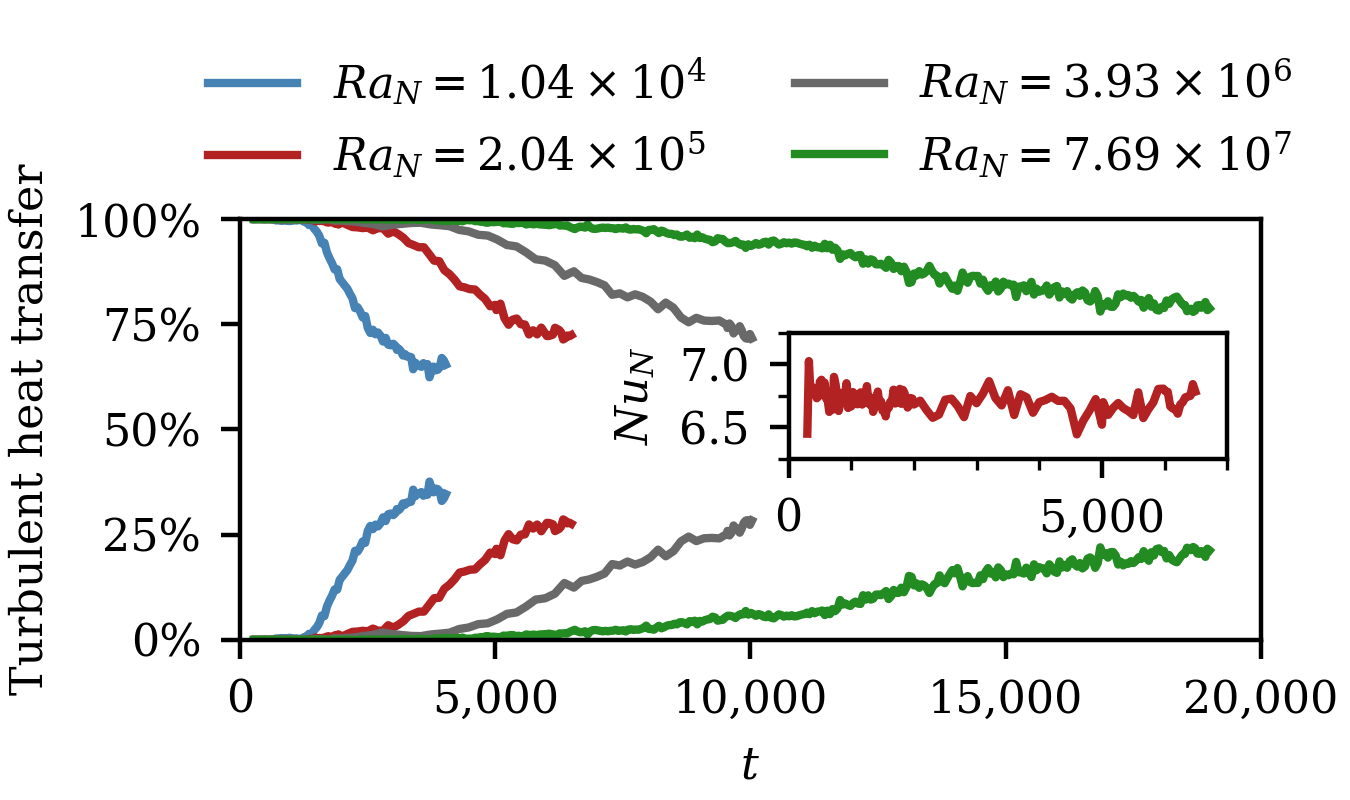}
\caption{Supergranule contribution to heat transfer. The time evolution of the turbulent convective heat flux across the midplane is compared at different ${\rm Ra}_{N}$. We show the time evolution of the relative fraction of transport due to the supergranule (bottom curves) and due to the rest of the turbulent structures (top curves). The inset demonstrates for the run Nfs2 that the Nusselt number that comprises both fractions remains on average constant over the whole time. Equally colored curves correspond to the same Rayleigh number as indicated in the legend.}
\label{fig4}
\end{center}
\end{figure}
%------------------------------------------------------------------------------------------

Despite this strong aggregation in physical and Fourier space, the global heat transfer and thus the Nusselt number ${\rm Nu}_N$ remain on average constant over the whole time period of the simulation in all runs as shown in the inset of Fig. \ref{fig4} for one example. Figure \ref{fig4} demonstrates, however, that the slow formation of the supergranule shifts the relative fraction of convective heat flux in the course of the evolution. We apply a filter in Fourier space and define
\begin{equation}
J_{k_{\rm min}}(z=0.5,t)=\frac{\langle {\cal F}^{-1}[u_zT (|{\bm k}|=k_{\rm min},t)]\rangle_A}{\langle u_z T\rangle_A}
\end{equation} 
and the rest $1 - J_{k_{\rm min}}(z=0.5,t)$ with $k_{\rm min}=2\pi/\Gamma$ as stated already above. It can be seen in all cases how the contribution of the supergranule structure gains importance for later times and reaches a statistically steady transport regime which is indicated in all runs (except Nfs3 which would eventually also saturate if run longer). The share of the structure to the global transport drops from nearly 40\% for the lowest to about 25\% for the highest ${\rm Ra}$ which underlines its relevance for the turbulent heat transfer across the convection layer. 

\section{Leading Lyapunov vector analysis}
A better understanding of the physical mechanisms of the aggregation can be obtained by applying a technique that is well-established in dynamical systems -- the Lyapunov analysis \cite{Pikovsky2016}. In this framework, the evolution of the turbulent convection flow corresponds to a trajectory in a very high-dimensional phase or state space (strictly speaking this state space is infinite-dimensional). The state of the fluid flow at time $t$ is given by a column vector ${\bm y}(t)=({\bm u}({\bm x}_k,t), T({\bm x}_k,t))$ which has $k=1, \dots, N_e N^3$ entries and thus $N_{\rm dof}=4 N_e N^3\gg1$ is the number of degrees of freedom in the present numerical model (see also Table I). The compact form of our Boussinesq flow is thus
\begin{equation}
\dot{\bm y}(t)={\bm F}({\bm y}, t)\,.
\label{ode}
\end{equation}
The sensitivity of the trajectory with respect to infinitesimal perturbations or in other words the tendency to develop new instabilities out of the present (fully turbulent) state can be probed by the strength of exponential separation of two initially very close trajectories, ${\bm y}(t)$ and ${\bm y}(t) + \delta {\bm y}(t)$, of the present turbulent flow. The corresponding linearized equations to \eqref{ode} are given by $\delta \dot{\bm y}(t)={\bm J}({\bm y}(t)) \delta {\bm y}(t)$ with the Jacobian ${\bm J}=\partial{\bm F}/\partial{\bm y}$. Here, $\delta{\bm y}(t)=(\delta{\bm u}({\bm x}_k,t), \delta T({\bm x}_k,t))$ is the infinitesimal perturbation field to the original trajectory of the flow. In detail, this gives the following set of linearized Boussinesq equations for our study
\begin{align}
\nabla \cdot \delta\bm{u} & = 0\,, \label{ce1}\\
\frac{\partial \delta \bm{u}}{\partial t} + ( \bm{u} \cdot \nabla ) \delta \bm{u} + (\delta \bm{u} \cdot \nabla ) \bm{u}& = - \nabla \delta p  \nonumber\\
                                                                                                                                      + \sqrt{\frac{\rm Pr}{{\rm Ra}_{D,N}}} & \nabla^{2} \delta \bm{u} + \delta T \bm{e}_{z} \,,\\
\frac{\partial \delta T}{\partial t} + ( \bm{u} \cdot \nabla ) \delta T + ( \delta\bm{u} \cdot \nabla ) T& = \frac{1}{\sqrt{{\rm Ra}_{D,N} {\rm Pr}}} \nabla^{2} \delta T \label{te1}\,,
\end{align}
where the pressure perturbation field $\delta p$ is determined by $\delta \bm u$ with a Poisson equation similar to the original incompressible case \eqref{ce}--\eqref{te}. The determination of the spectrum of the first $n$ of the total of $N_{\rm dof}$ Lyapunov exponents, $\lambda_1\ge \lambda_2\ge \dots \lambda_n$ and their corresponding Lyapunov vector fields requires the simultaneous numerical solution of $n$ versions of \eqref{ce1}--\eqref{te1} with different initial perturbations $\delta {\bm y}_1(0), \delta {\bm y}_2(0), \dots, \delta {\bm y}_n(0)$ in combination with the original equations \eqref{ce}--\eqref{te}. The computational complexity of this task in an extended turbulent convection flow limits us here to the leading Lyapunov exponent $\lambda_1(t)$ and the corresponding leading Lyapunov vector field which encodes the locations in the fluid volume and associated scales of the flow patterns that become unstable first. As pointed out by Levanger et al. \cite{Levanger2019}, the local magnitude of the components of the Lyapunov vector indicate the sensitivity of these local regions with respect to perturbations. It thus contains the essential information that we need to explain the aggregation. The leading Lyapunov exponent $\lambda_1(t)$ is given by 
%------------------------------------------------------------------------------------------
\begin{equation}
\lambda_{1}(t) = \frac{d}{dt} \log \left( \frac{\| \delta {\bm y}_1 (t) \|}{\| \delta {\bm y}_1 (0) \|} \right)\,,
\label{eq:lyapunov_exponent}
\end{equation} 
%------------------------------------------------------------------------------------------
with a norm that is determined by
\begin{equation}
\|\delta {\bm y}_1(t)\|=\sqrt{\frac{1}{V}\int_V (\delta{\bm u}_1(t)^2+\delta T_1(t)^2) dV}\,.
\end{equation}
%------------------------------------------------------------------------------------------
\begin{figure}
\begin{center}
\includegraphics[scale=1.00]{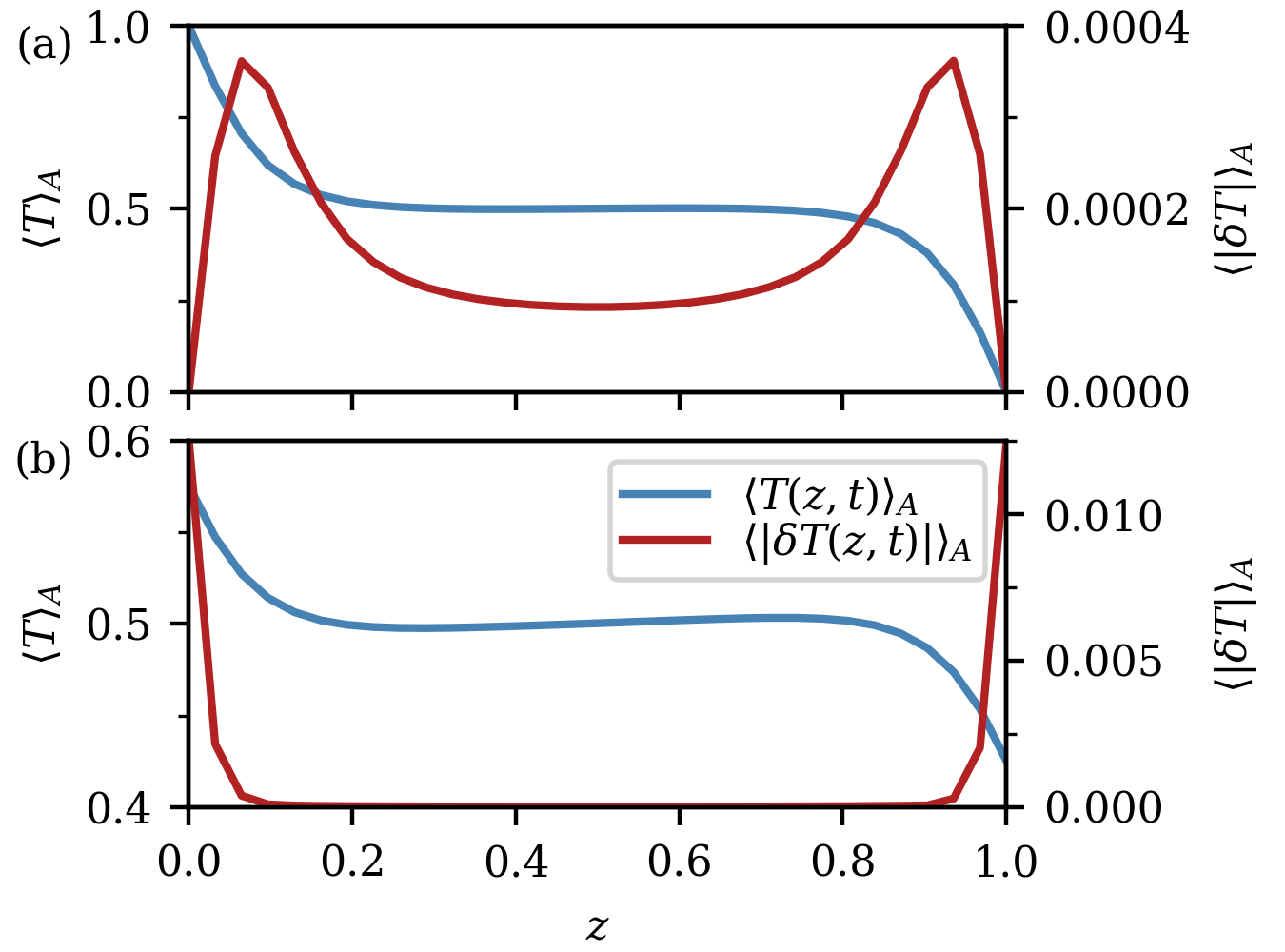}
\caption{Comparison of mean profiles. The vertical plane-averaged profiles of the temperature field $T({\bm x},t)$ and the absolute value of the temperature component of the leading Lyapunov vector field $|\delta T({\bm x},t)|$ are compared for (a) simulation run Dfs2 at $t=750$ and (b) Nfs2 at $t=6,500$. The local maxima in panel (a) are at $z_0\approx 2/3\delta_T$ and $1-2/3\delta_T$.}
\label{fig5}
\end{center}
\end{figure}
%------------------------------------------------------------------------------------------
\begin{figure*}
\begin{center}
\includegraphics[scale=1.00]{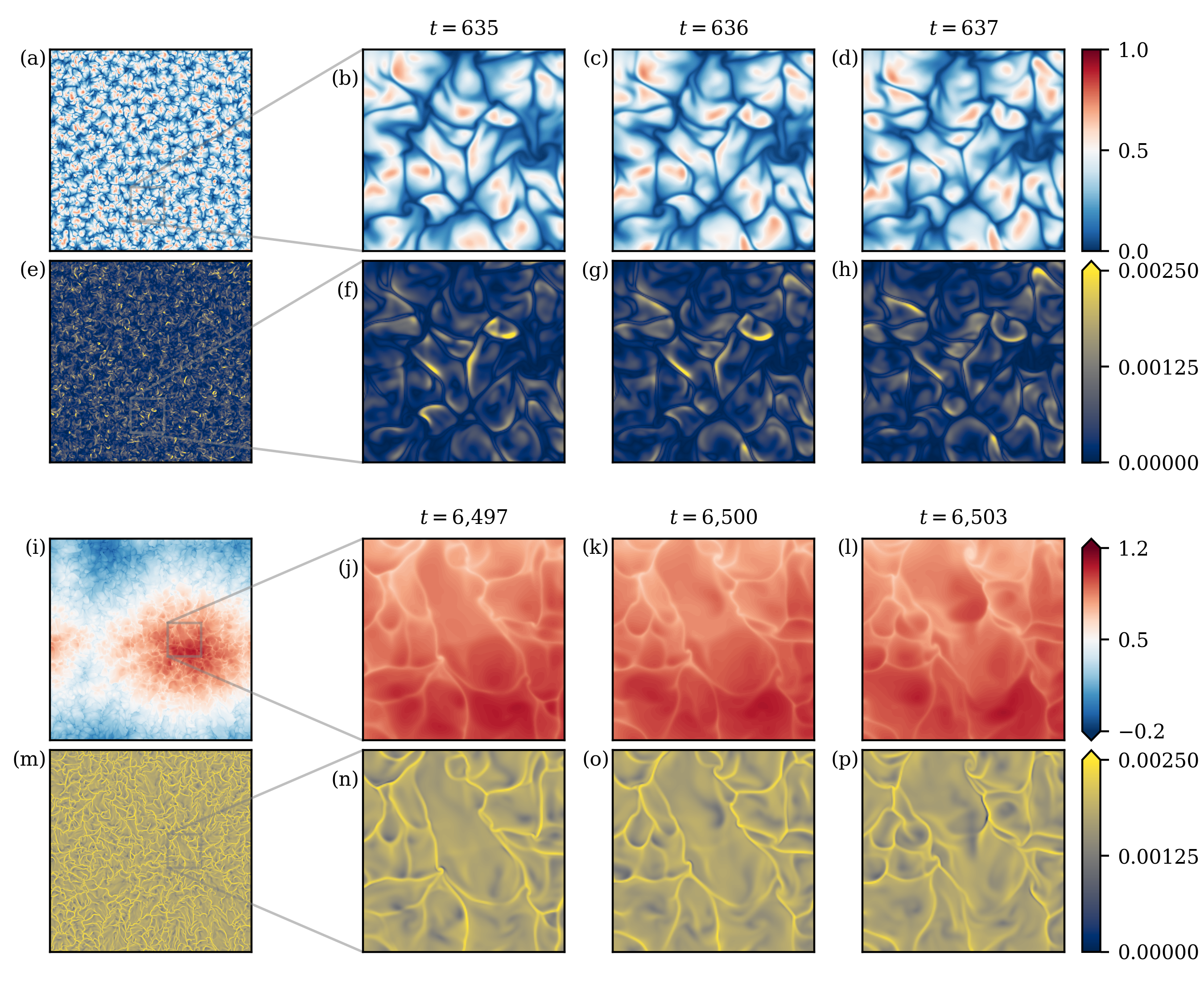}
\caption{Lyapunov vector fields for Dirichlet and Neumann boundary conditions. The time evolution of patterns in the temperature field $T(x,y,z_0,t)$ (a--d,i--l) and the leading Lyapunov vector temperature perturbation field $| \delta T(x,y,z_0,t) |$ (e--h,m--p) is visualised. We show data for run Dfs2 at $z_0\approx 1-2/3\delta_T\approx 0.94$ (a--h) and run Nfs2 at $z_0=1-\delta_T/2$ (i--p). The value of $z_0\approx 0.94$ in Dfs2 corresponds with one of the two maxima in Fig. \ref{fig5}(a). Panels (a,i) show $T$ and panels (e,m) $\delta T$ over the whole cross-section of size $L \times L = 60 \times 60$. The remaining subplots enlarge a highlighted region of interest of size $10 \times 10$. The times of the corresponding snapshots are indicated above. See also appendix C for the same analysis of Dfs2 in the midplane.}
\label{fig6}
\end{center}
\end{figure*}
%------------------------------------------------------------------------------------------
The resulting spatial ridge patterns of the components of the leading Lyapunov vector reflect a critical wavelength (or scale) across which the instability is triggered. We have tested our algorithm for a RBC flow in the weakly nonlinear regime at ${\rm Ra}=5\times 10^3\gtrsim {\rm Ra}_c$ where this technique has been established \cite{Egolf2000,Jayaraman2006,Levanger2019,Scheel2006}, see appendix C.
%------------------------------------------------------------------------------------------
\begin{figure*}
\begin{center}
\includegraphics[scale=1.00]{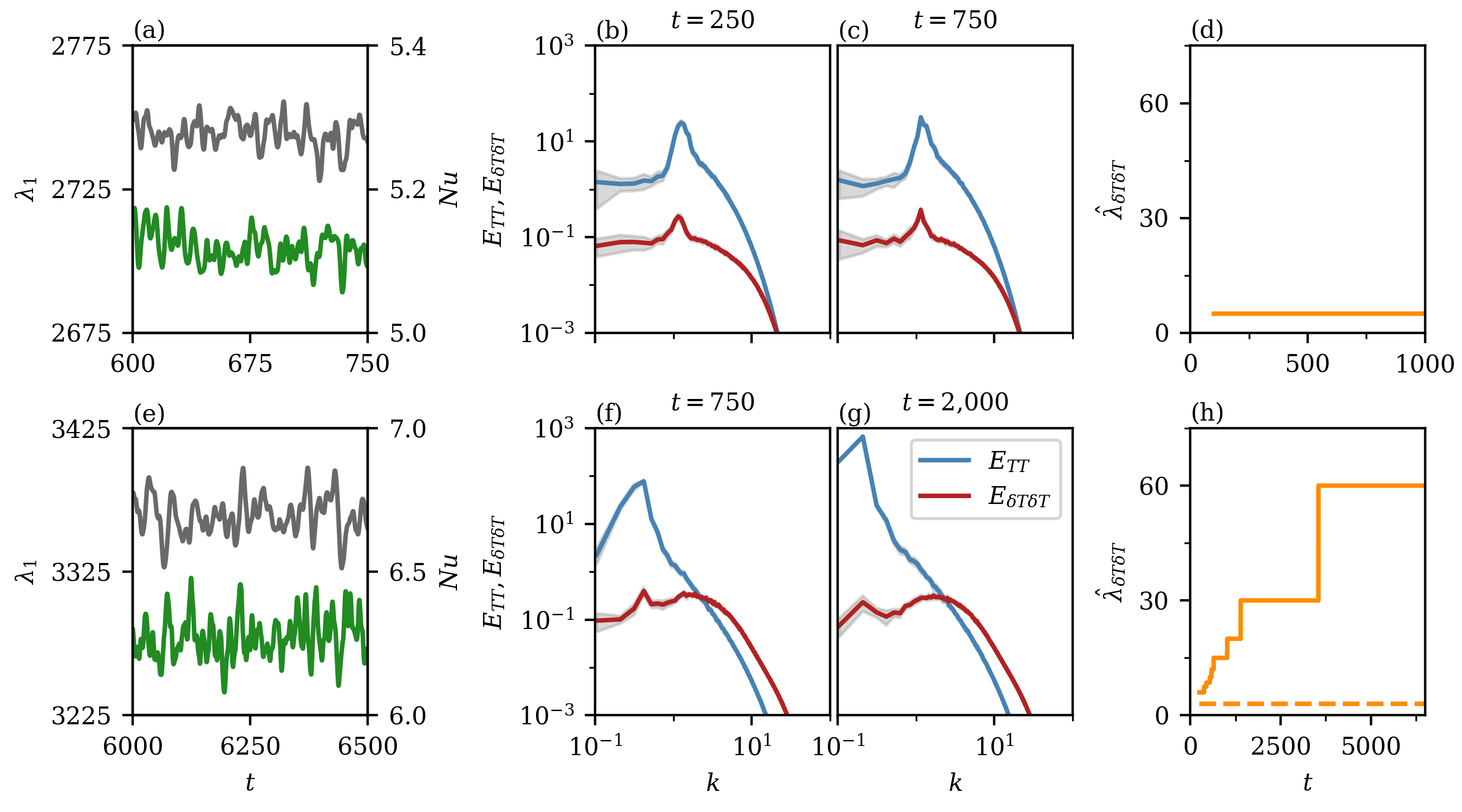}
\caption{Unstable scale extraction from leading Lyapunov vector field. Time series of the leading Lyapunov exponent $\lambda_{1}$ (green) and the Nusselt number $Nu$ (grey) versus time for run Dfs2 in (a) and for corresponding run Nfs2 in (e). The Fourier spectra of temperature $E_{TT}$ and temperature component of the leading Lyapunov vector $E_{\delta T \delta T}$ are analysed for run Dfs2 at $z_0=1-2/3\delta_T\approx 0.94$ in panels (b,c) and for Nfs2 at $z_0=1-\delta_T/2$ in panels (f,g). Corresponding times for the analysis are indicated at the top of each panel. Time-averaged spectra are taken over 50 subsequent snapshots centered around the one with time indicated at the top. The shaded regions represent fluctuations with respect to the standard deviation. Also, $E_{\delta T \delta T}$ is multiplied by $10^4$ in (b,c) and $10^5$ in (f,g) for a better comparison. Panels (d) and (h) display the temporal evolution of the wavelength $\hat{\lambda}_{\delta T\delta T}$  that corresponds to the local maximum in the spectrum $E_{\delta T\delta T}(k)$ at a wavenumber $k\lesssim 1$ (solid lines). This wavenumber drifts towards $L$ in Nfs2 while it remains at $\hat{\lambda}_{\delta T\delta T}\approx 4$ in Dfs2. The dashed line in panel (h) stands for the second local maximum at a smaller scale $\Lambda\approx 3$.}
\label{fig7}
\end{center}
\end{figure*}
%------------------------------------------------------------------------------------------

Figure \ref{fig5} compares the mean vertical profiles of the temperature field, $\langle T(z,t)\rangle_A$, and the absolute value of temperature component of the (renormalized) leading Lyapunov vector field $\delta \tilde{\bm y}_1$ which is denoted as $\langle |\delta T(z,t)|\rangle_A$ for cases with Dirichlet and Neumann boundary conditions. For simplicity, we denoted the fourth component of $\delta\tilde{\bm y}_1$ again by $\delta T$. We take the absolute value as we are interested in the magnitude only. Here $\langle\cdot\rangle_A$ denotes an average over the whole cross section plane $\Gamma^2$. While the mean temperature profile of Dfs2 has a zero slope in the bulk, the one for Nfs2 is slightly stably stratified. Such a subadiabaticity of the temperature is considered as a possible origin of the emergence of supergranules in the Sun \cite{Brandenburg2016,Cossette2016}. The mean profiles of the Lyapunov vector component are found to differ qualitatively as reported in the same figure. In case of Dirichlet boundary conditions, the most sensitive region is located at the top of the thermal boundary layer where plume mixing starts. This is in contrast to Neumann boundary condition case, where the bottom and top planes are found to be by far most susceptible with respect to small perturbations.

We select the plane that corresponds to one of the two local maxima in Fig. \ref{fig5}(a) at $z_0\approx 0.94$, i.e., where the instability has the largest magnitude, and show in Fig.~\ref{fig6} (a-h) that local instabilities and thus the maxima of $|\delta T(x,y,z_0,t)|$ are found close to (but not at) the downflow regions in Dfs2. They also remain connected to local creation or annihilation of defects of the flow patterns for the turbulent and fully time-dependent Dirichlet boundary case. Despite operating in the turbulent regime of the flow, these instabilities are thus very similar to what is found for the weakly nonlinear regime. We display therefore a magnification of a short dynamical sequence in panels (b-d, f-h).
In the corresponding turbulent Neumann boundary case Nfs2 in Fig. \ref{fig6} (i-p) at ${\rm Ra}_{N} = 2.04 \times 10^{5}$, the structure of the temperature perturbation $|\delta T(x,y,z_0,t)|$ is different. The plane that was now selected is taken close to the top wall at $z_0=1-\delta_T/2$  in correspondence with Fig. \ref{fig5}(b). One observes a connected pattern of high-amplitude ridges of $\delta T$ with a coarser spacing indicating a larger scale of instability. It is also observed now that the locally most unstable regions coincide with the downflow regions thus stabilizing the bulk regions (see also \cite{Brandenburg2016,Spruit1996} for similar mechanisms in solar case).

Figure \ref{fig7} provides additional results on the instabilities as well as on the corresponding scales. First, we show time series of $\lambda_1(t)$ and ${\rm Nu}(t)$ for Dfs2 (a) and Nfs2 (e). Both values vary with respect to time about the mean values which is typical for the turbulent flow case. Furthermore, Fourier spectra of the temperature, $E_{TT}(k)$, and the temperature component of the leading Lyapunov vector,  $E_{\delta T\delta T}(k)$, are shown. Data are taken in planes close to the top where maximum magnitudes of $\delta T$ are found.  The Dirichlet case displays a local maximum at $k=k_m$ that coincides for both spectra (see panels (b,c)). This local spectral peak corresponds to a characteristic wavelength which is given by 
\begin{equation}
\hat{\lambda}_{\delta T\delta T}=\frac{2\pi}{k_m}\,,  
\end{equation}
which remains unchanged in time at a value of $\hat{\lambda}_{\delta T\delta T}\approx 4$ and thus corresponds to the characteristic extension of the turbulent superstructures which have been discussed in Pandey et al. \cite{Pandey2018} and Fonda et al. \cite{Fonda2019}. 
The results differ for the Neumann case as displayed in panels (f-h) of the figure. The spectrum $E_{\delta T\delta T}(k)$ has a large-wavenumber bump at $k_m\approx 2$ which is indicated by the dashed line in Fig. \ref{fig7}(h). These instabilities correspond to the fine granule patterns which are for example seen in Fig. \ref{fig2}(b).  While this local maximum remains unchanged in time, a second one moves gradually towards larger wavelengths, see again Fig. \ref{fig7}(h). This suggests that the turbulent flow develops instabilities at an increasingly larger wavelength. The process is ceased when the system size is reached and the nonlinear process of supergranule formation is completed. We stress once more that this behavior is fundamentally different to the case with Dirichlet boundary conditions for the temperature field.

\section{Discussion and perspective} 
Our main motivation was to demonstrate the gradual formation of a salient large-scale convection pattern on a time scale larger than the vertical diffusion time $\tau_v=\sqrt{\rm Pr\,Ra}$ and that eventually fills the whole convection domain in a Rayleigh-B\'{e}nard convection setup. Following solar convection, this structure is termed {\em supergranule}.  We showed that this formation proceeds only in case of constant heat flux boundary conditions (also known as Neumann conditions) at the top and bottom planes of the layer, independently of no-slip or free-slip boundary conditions for the velocity field. Surprisingly the supergranule pair is still observed when the flow is in the state of fully developed turbulence as being the case, at least for runs Nfs3 and Nfs4. We mention here simulations of compressible photospheric convection (with similar boundary conditions) by Rincon et al. \cite{Rincon2005} at $\Gamma=42$. The authors report the formation of a dominant convective mode and conclude that the simulations could not be run long enough to study a  further aggregation. Our results confirm that long simulation times are necessary and demonstrate that this dominant convection mode is eventually a supergranule pair which can be seen even in the simpler (incompressible) RBC setup.  

As discussed, the critical mode at onset of Rayleigh-B\'{e}nard convection in an infinitely extended layer with Neumann conditions at the top and bottom is $k_c=0$. This implies that a pair of counter-rotating convection cells fills a domain with a finite periodicity length $L$ at onset, a behavior which is found in this work to persist to ${\rm Ra}_{N}\gg {\rm Ra}_{N,c}$. We confirm the behavior by detecting the accumulation of kinetic energy and thermal variance in the four next neighboring Fourier modes to the (critical) zero mode with wavelength $L$. The determination of the leading Lyapunov vector field and the subsequent spectral analysis of its temperature component, demonstrates clearly that the flow structures at a given scale give rise to an instability at a next bigger wavelength and thus to a spatially larger new flow structure. This inverse cascade continues until the horizontal periodicity length is reached for the present setup. In the solar case a further physical process will limit the supergranule size. 

We thus demonstrate that the structure formation mechanism, which was described in Chapman and Proctor \cite{Chapman1980} above the onset of convection in the weakly nonlinear regime, persists far into the turbulent range. A possible next step would be to derive effective amplitude equations, now for the perturbations about a fully turbulent state. This will include turbulent closures and certainly requires simplifying assumptions, but could be done along lines of a very recent work by Ibbeken et al. \cite{Ibbeken2019}. Our Lyapunov vector analysis answers furthermore a question left open in \cite{Pandey2018}: the generation of turbulent superstructures in the Dirichlet case is a local pattern instability with a scale of the size of a pair of counter-rotating mean circulation rolls, here $\hat{\lambda}_{\delta T\delta T}\approx 4$. In contrast,  the Neumann case proceeds slowly to a global wavelength instability by a cascading process with $\hat{\lambda}_{\delta T\delta T}\to L$.    

Finally, we return to the initial example of solar convection where the fixed heat flux at the top is connected with the well-known solar luminosity $L_{\odot}$. This flux is the main driver of convection and thus the formation of granules and supergranules in the upper convection zone. Our study showed that already these boundary conditions alone generate a large-scale convection roll pair, i.e., without additional magnetic fields, changes in chemical composition and the strong compressibility effects. As the typical scale ratio $\ell_{\rm SG}/\ell_{\rm G}\approx 30$ of the solar convection case is equal to the diameter ratio of our supergranule to granule roll for the prescribed layer extension, namely $(\Lambda/2)/H\approx 30$, we want to compare now characteristic velocities and evolution times of Nfs4 with the solar data given in the introduction.  We thus decompose ${\bm u}(\bm x, t)={\bm U}({\bm x})+{\bm u}^{\prime}({\bm x},t)$  with ${\bm U}({\bm x})=\langle {\bm u}({\bm x},t)\rangle_t$. The ratio of the corresponding root mean square velocities $u^{\prime}_{\rm rms}/U_{\rm rms}\approx 5.8$ comes close to the velocity ratio of $v_{\rm G}/v_{\rm SG}\approx 6$ \cite{Hathaway2013}. When the lifetime of a granule is estimated by the mean turnover time of a Lagrangian tracer across the layer with a value of $t_{\rm to}\approx t_{u^{\prime}}\sim 20$ \cite{Schneide2018}, one arrives at $t_U/t_{u^{\prime}} \sim 10^4/20\sim 500$  which is at least of the same order of magnitude as  $\tau_{\rm SG}/\tau_{\rm G}\approx 144$ \cite{Hathaway2013}. 

Clearly, this approximate agreement should be taken with caution as the solar convection zone contains a much more complicated physics at a much larger Rayleigh number and an extremely small Prandtl number, ${\rm Pr}\lesssim 10^{-6}$ \cite{Schumacher2020}. Nevertheless, our simple convection model might still turn out to be fruitful to better interpret the solar observations as we were able to reveal a basic instability mechanism in this class of turbulent flows that leads to a large-scale flow structure. It is thus also a good starting point for a step-by-step increase of complexity towards the solar case that can test how the supergranule formation is affected by an inclusion of further physical processes. A promising extension would be to include constant rotation about the vertical axis into the present model as an additional process that stops the horizontal growth of the supergranules before reaching domain size $L$ (see again ref. \cite{Featherstone2016}). These studies started very recently and will be reported elsewhere.

\section*{Acknowledgements}
The work of PPV and JDS is supported by the Deutsche Forschungsgemeinschaft within the Priority Programme DFG-SPP 1881 on Turbulent Superstructures. 
The authors gratefully acknowledge the Gauss Centre for Supercomputing e.V. (www.gauss-centre.eu) for funding this project by providing computing time on the GCS Supercomputer SUPERMUC-NG at Leibniz Supercomputing Centre (www.lrz.de) and through the John von Neumann Institute for Computing (NIC) on the GCS Supercomputer JUWELS at J\"ulich Supercomputing Centre (JSC). We thank Tao Wang Kwan for his help on the Lyapunov solver and Vincent B\"oning, Charles R. Doering, David Goluskin, Ambrish Pandey and Katepalli R. Sreenivasan for helpful discussions.

\section*{Appendix A -- Spectral resolution tests}
\begin{figure}
\begin{center}
\includegraphics[scale=0.99]{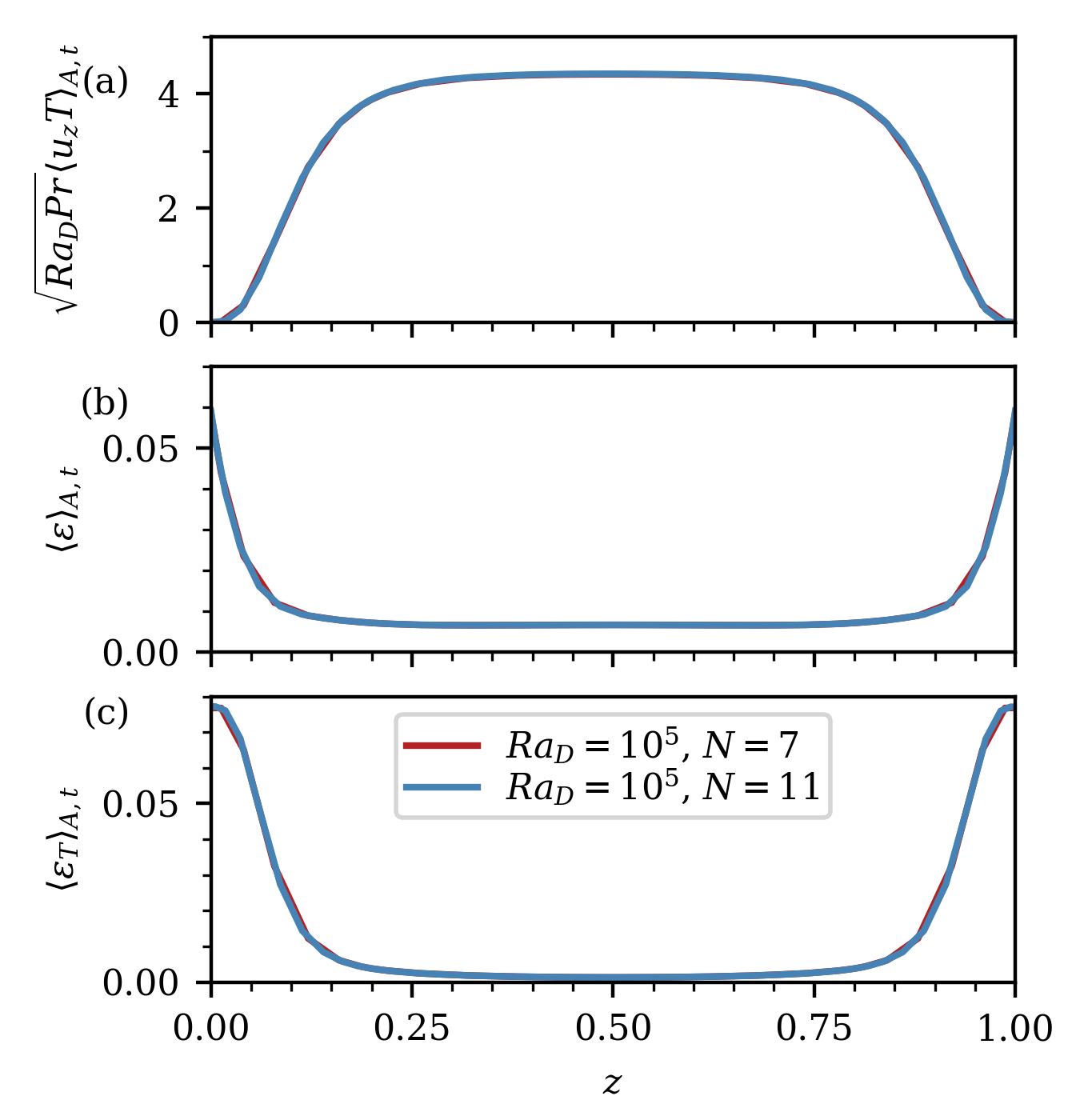}
\caption{Spectral resolution study with respect to the polynomial order $N$ on each spectral element. Vertical profiles of plane-time-averaged quantities of one simulation setup with ${\rm Ra}_{D} = 10^{5}$, ${\rm Pr} = 1$ and no-slip boundary conditions at the top and bottom are shown. We use the same spectral element mesh as for simulations Dfs2, Nfs1, Nfs2 with $N_{e} = 160,000$ and plot the mean profiles of the convective heat current $\sqrt{{\rm Ra}_{D} {\rm Pr}} \langle u_{z} T \rangle_{A, t}$ (a), the kinetic energy dissipation rate $\langle \epsilon \rangle_{A, t}$ (b), and the thermal dissipation rate $\langle \epsilon_{T} \rangle_{A, t}$ (c).}
\label{fig_SI_1}
\end{center}
\end{figure}

In spectral element methods, the resolution depends on two factors: the total number of spectral elements ($N_e$) and the polynomial order of the spectral expansion on each element and in each spatial direction ($N$). The method belongs to the bigger class of exponentially fast converging $hp$-FEM (FEM=finite element method) where equations are solved on elements that fill the volume by a piecewise polynomial approximation of the solutions. One can vary the size of the element $h$ (here $N_e$) or the polynomial degree $p$ (here $N$).  In our flow, the polynomial order and spectral element number has to be chosen properly such that the steep gradients near the top and bottom walls and the Kolmogorov scale $\eta_K$ can be resolved sufficiently. Sufficient resolution is established once, $dz (z)/\langle \eta_{K}(z) \rangle_{A, t} \lesssim \pi/2$ for ${\rm Pr} \gtrsim 1$. Here $dz(z)$ is the vertical spectral element extension. This criterion was suggested and tested in ref. \cite{Scheel2013}. To show the convergence of our results, we perform first a resolution test with respect to two different polynomial orders $N$, as shown in Fig. \ref{fig_SI_1}. The production run setup is at $N_{e} = 160,000$ and $N = 11$ (case Dns2). It can be seen that the same results can be achieved already with lower polynomial order of $N = 7$ and that the curves for both $N$ collapse. We conclude that the spectral resolution with $N=11$ is sufficient.
%------------------------------------------------------------------------------------------
%------------------------------------------------------------------------------------------

\begin{figure}
\begin{center}
\includegraphics[scale=0.99]{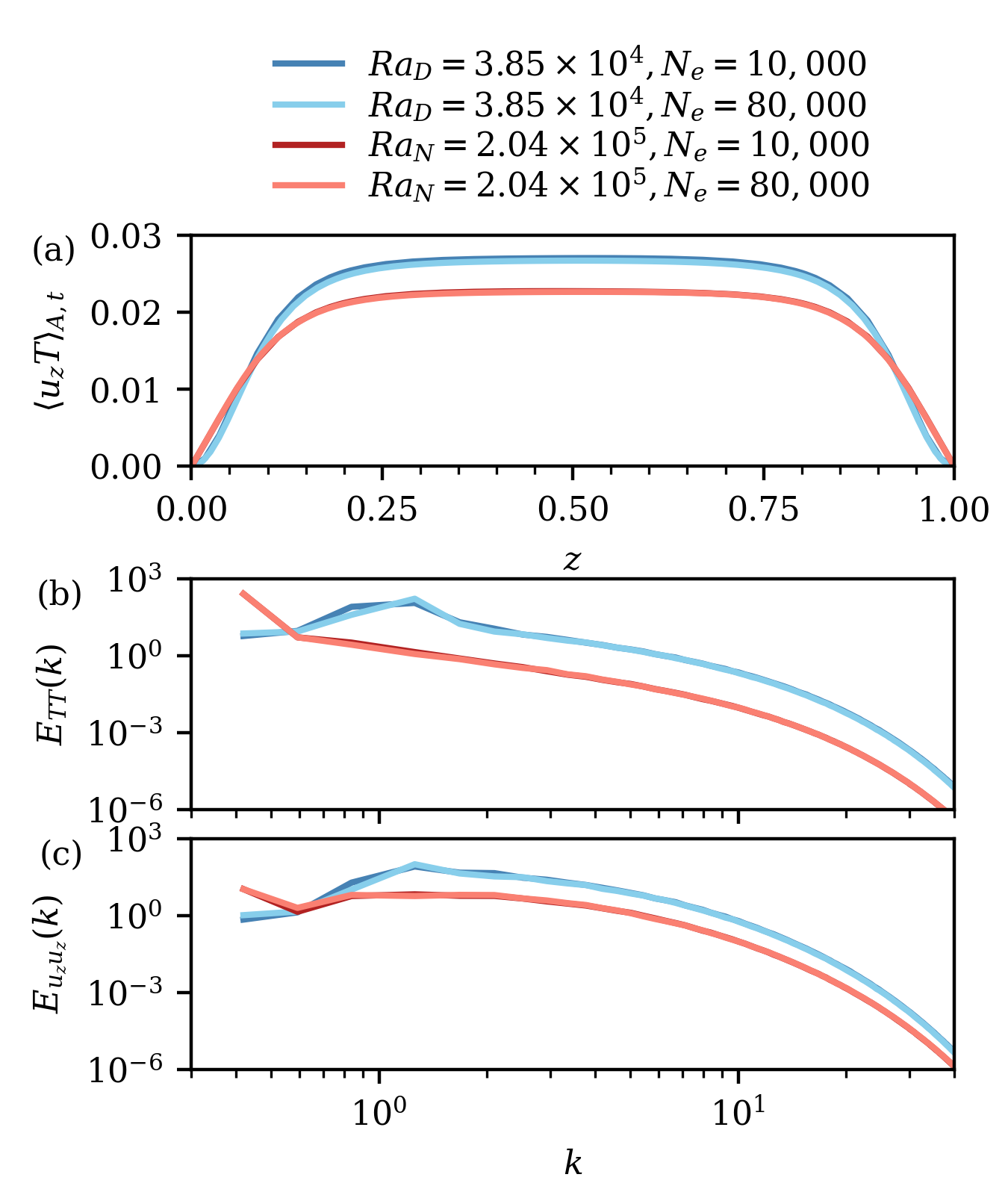}
\caption{Spectral resolution study with respect to the number of spectral elements $N_{e}$ at fixed polynomial order $N=11$. A comparison of Dirichlet and Neumann boundary conditions is shown as indicated in the legend at the top. The aspect ratio is $\Gamma = 15$, the Prandtl number is fixed at ${\rm Pr} = 1$, and free-slip boundary conditions are applied at the plates for the velocity field in all simulations. The lateral boundary conditions are periodic. The Rayleigh number for the case of constant flux ${\rm Ra}_{N}$ is agrees with case  Nfs2, whereas ${\rm Ra}_{D}$ is equal to that of case Dfs2 (see Table 1). Profiles in panel (a) show convective heat transfer $\langle u_{z} T \rangle_{A, t}$. The data for the Neumann case have been multiplied by a factor of 10 for better visibility. Plots (b,c) display the time- and azimuthally averaged Fourier spectra of temperature and vertical velocity component in the midplane. The time average is performed over a time span of 200 convective free-fall times in all cases.}
\label{fig_SI_3}
\end{center}
\end{figure}

Furthermore, we demonstrate that the gradual supergranule formation is not a resolution effect. This is done in a smaller cell of aspect ratio $\Gamma = 15$ to accelerate the formation process. We use the parameters of simulation run Nfs2 with ${\rm Ra}_{N} = 2.04 \times 10^{5}$ and the corresponding run Dfs2 with ${\rm Ra}_{D} = 3.85 \times 10^{4}$. We apply the same spectral element resolution as in the main text, which translates to $N_{e} = 10,000$. In the corresponding comparison runs, we double the number of elements in each space direction which leads to $N_{e} = 80,000$. The supergranule evolves in the long-term dynamics in case of Neumann boundary conditions, while there is no such effect for Dirichlet boundary conditions. The results are summarized in Fig. \ref{fig_SI_3}. In panel (a), it is seen that the convective heat flux profiles collapse on each other for both pairs of runs.  The Fourier spectra in panels (b,c), which are taken for the Dirichlet run from 750 to 950 $t_f$ and for the Neumann run from 2,500 to 2,700 $t_f$, display the aggregation in the latter case which agrees very well with the results for $\Gamma =60$.  In the Neumann boundary case, the supergranule is already fully developed in both runs. Note that in most panels the curves collapse onto each other.
%------------------------------------------------------------------------------------------
%------------------------------------------------------------------------------------------

\section*{Appendix B -- Velocity boundary conditions}
\label{sec:comparison_BCs}
\begin{figure*}[b]
\begin{center}
\includegraphics[scale=0.95]{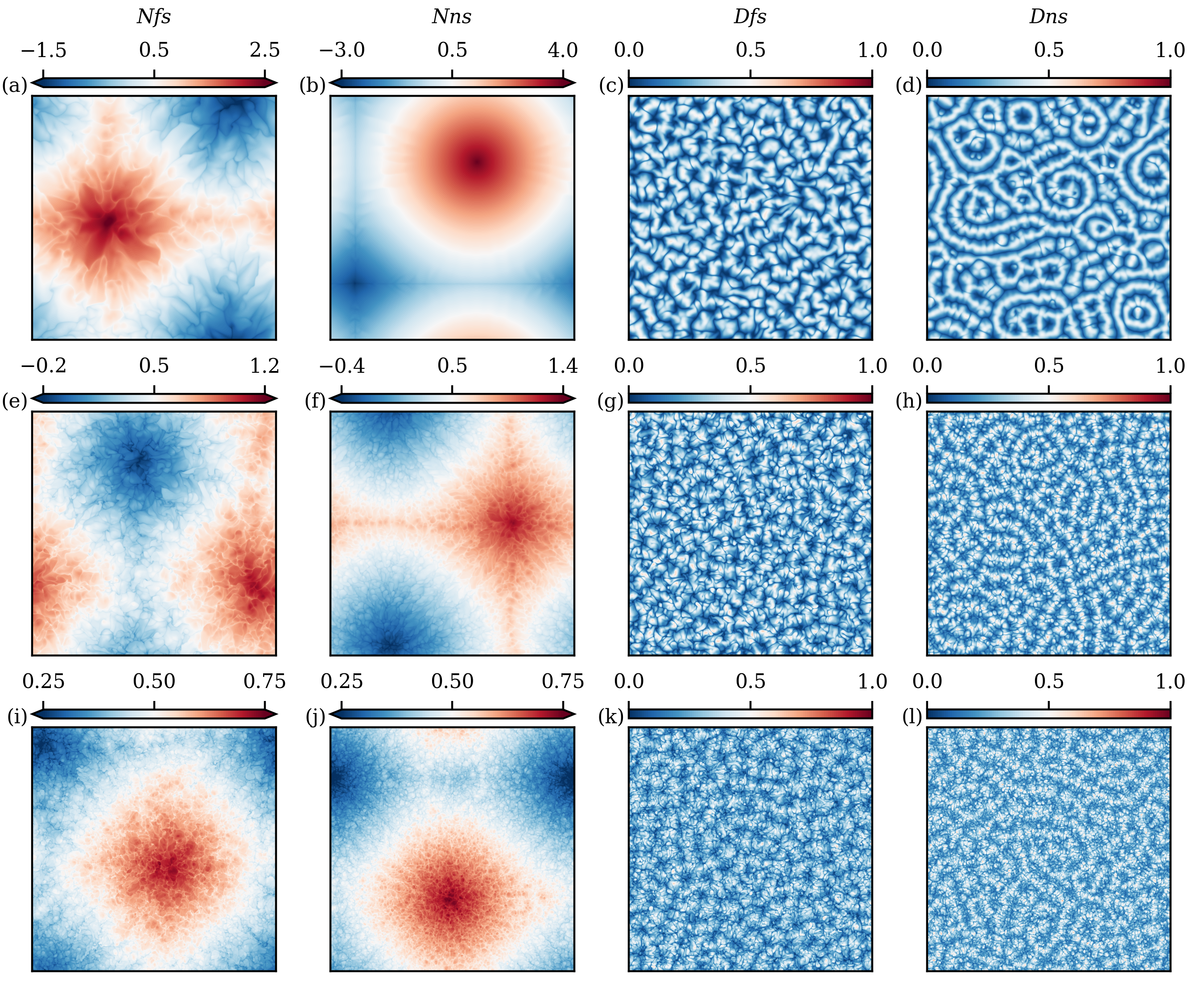}
\caption{Comparison of the impact of Neumann (N) and Dirichlet (D), as well as free-slip (fs) and no-slip (ns) boundary conditions on the instant temperature field $T(x,y, z=1-\delta_{T}/2,t_0)$ in the upper boundary layer. The specific combination of thermal and velocity boundary conditions is given at the top of each column. Lateral boundaries obey periodicity and ${\rm Pr}=1$ for all simulations. The Rayleigh number increases from top to bottom, i.e. we plot case (Nfs1, Nfs2, Nfs3) in panels (a, e, i) -- other columns similarly. The corresponding Rayleigh numbers in the figure are given in the text of appendix B.}
\label{fig_SI_2}
\end{center}
\end{figure*}

We compare in Fig.~\ref{fig_SI_2} four different combinations of temperature and velocity field boundary conditions for snapshots of the temperature field at $z = 1 - \delta_{T}/2$ and at a late stage of the dynamical evolution. The panels in the leftmost column coincide with those in Fig.~\ref{fig1}. It can be seen that the instantaneous temperature patterns have  very different characteristics for the Neumann and Dirichlet cases. It is also seen that the change of temperature boundary conditions is the essential one (and not the change of the velocity boundary conditions) that leads to the supergranule. All fields are visualized for the whole cross-section of size $L \times L = 60 \times 60$. 

We start with ${\rm Ra}_D=10^4, 10^5$, and  $10^6$ for runs Dns1, Dns2, and Dns3, respectively. In order to have the same distance from the onset of convection, we take ${\rm Ra}_D=3.85\times 10^3, 3.85\times 10^4$, and $3.85\times 10^5$ for runs Dfs1, Dfs2, and Dfs3, respectively. The corresponding two series with Neumann boundary conditions follow by eq. \eqref{comparison}. Thus  ${\rm Ra}_N=2.23\times 10^4, 4.34\times 10^5$, and  $8.31\times 10^6$ for runs Nns1, Nns2, and Nns3, respectively. The corresponding Rayleigh numbers for Nfs1, Nfs2, and Nfs3 are listed in Table 1.
%------------------------------------------------------------------------------------------
%------------------------------------------------------------------------------------------ 

\section*{Appendix C -- Lyapunov vector determination}
\begin{figure*}[b]
\begin{center}
\includegraphics[scale=0.95]{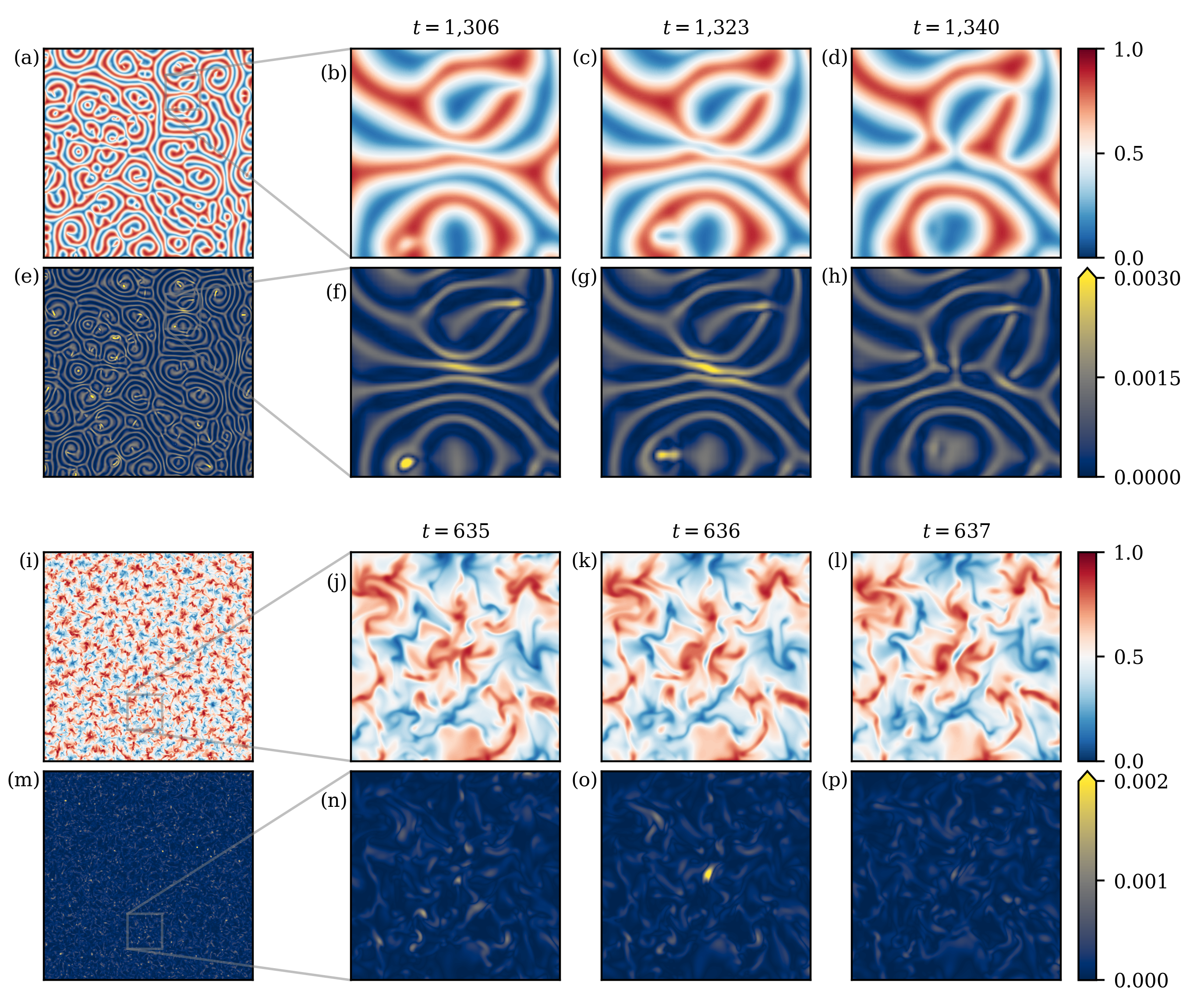}
\caption{Lyapunov analysis in an RBC flow in the spiral defect chaos (a-h) and turbulent flow regimes (i-p). The time evolution of the temperature field $T(x,y,z=0.5,t)$, see panels (a-d) and (i-l), and the temperature perturbation field of the leading Lyapunov vector $| \delta T(x,y,z=0.5,t) |$, see panels (e-h) and (m-p). Dirichlet and no-slip boundary conditions are shown here. Lateral boundaries are periodic, the aspect ratio $\Gamma = 60$, and the Prandtl number ${\rm Pr} = 1$. The Rayleigh numbers are ${\rm Ra}_D = 5\times 10^3$ and $3.85 \times 10^4$ (case Dfs2). Panels (a,e,i,m) show the whole plane $L \times L = 60 \times 60$, while the remaining panels enlarge a marked region of size $10 \times 10$. The time of the corresponding snapshots is indicated at the top of each column.}
\label{fig_SI_5}
\end{center}
\end{figure*}

We provide in Fig.~\ref{fig_SI_5} two series of snapshots that show the evolution of the temperature field $T(x,y,z,t)$ together with the temperature component of the corresponding leading Lyapunov vector, $\delta T(x,y,z,t)$, for a combination of Dirichlet and no-slip boundary conditions. Panels (a)--(h) are taken in the weakly nonlinear regime at ${\rm Ra}_{D} = 5.0\times 10^3$. This run is a test of our routine as it can be compared with results in the listed references, e.g. Egolf et al. \cite{Egolf2000} or Scheel et al. \cite{Scheel2006}. As in those references, the Lyapunov vector field highlights the regions of instability, where the defect formation is observable as a bright spot. The time-averaged leading Lyapunov exponent $\lambda_1=252\pm 1$. Panels (i)--(p) are for the turbulent flow case Dfs2 which is also discussed in the main text. Again, local maxima of the Lyapunov vector field correspond to a defect generation. The time-averaged leading Lyapunov exponent $\lambda_1=2703\pm 7$. All data which are shown in this figure are taken in the midplane $z=0.5$. The appearance of the localised defect is clearly detectable in the leading Lyapunov vector field for both cases, see panels (c,g,k,o) of the figure.
%------------------------------------------------------------------------------------------
%------------------------------------------------------------------------------------------

\end{document}